\documentclass[varenna]{cimento}

\usepackage{epsfig}
\usepackage{graphicx}
\usepackage{amsmath}
\usepackage{amsbsy}
\usepackage{amssymb}
\usepackage{amsfonts}
\usepackage{dsfont}

\title{Controlling hole spins in quantum dots and wells \label{chapt:QCS}}

\author{Stefano Chesi}

\institute{Beijing Computational Science Research Center, Beijing 100084, China}

\author{Xiaoya Judy Wang \atque W.~A.~Coish}

\institute{Department of Physics, McGill University, Montr\'eal, Qu\'ebec H3A 2T8, Canada}

\begin{document}

\maketitle

\begin{abstract}
We review recent theoretical results for hole spins influenced by spin-orbit coupling and Coulomb interaction in two-dimensional quantum wells as well as the decoherence of single hole spins in quantum dots due to hyperfine interaction with surrounding nuclear spins.  After reviewing the different forms of spin-orbit coupling that are relevant for electrons and heavy holes in III-V semiconductor quantum wells, we illustrate the combined effect of spin-orbit coupling and Coulomb interactions for hole systems on spin-dependent quasiparticle group velocities.  We further analyze spin-echo decay for a single hole spin in a nuclear-spin bath, demonstrating that this decoherence source can be controlled in these systems by entering a motional-averaging regime.  Throughout this review, we emphasize physical effects that are unique to hole spins (rather than electrons) in nanoscale systems. 
\end{abstract}

\section{Introduction}
Spin-related phenomena in semiconductor nanostructures have become increasingly important in many branches of condensed matter physics, including the study of spin currents \cite{zutic2004spintronics,fabian2007semiconductor} and spin-Hall effects \cite{Engel2007,vignale2010ten}, the search for Majorana fermions in solid-state systems \cite{lutchyn2010majorana,oreg2010helical,alicea2010majorana,alicea2011non,Mourik2012,Williams2012,Rokhinson2012}, topological insulators \cite{hasan2010colloquium,qi2011topological}, and applications of quantum information/computation \cite{loss1998quantum,awschalom2002semiconductor,Coish2006quantum,hanson2007spins,zak2010quantum,Ladd2010a}.  Improved understanding of these phenomena could lead to new technologies based on our ability to control electron spins. The ultimate demonstration of this control would be the precise manipulation of the quantum states of independent spins as well as interactions between them. Such a level of control for spins in semiconductor quantum dots may allow for a scalable implementation of quantum computing \cite{loss1998quantum,hanson2007spins,zak2010quantum,Coish2006quantum,Ladd2010a}. Additionally, semiconductor `spintronic' devices relying on spin-polarization and spin currents (rather than electric charges and charge transport) could lead to low-power high-density devices that would be unthinkable with conventional electronics \cite{awschalom2002semiconductor,zutic2004spintronics,fabian2007semiconductor}. A central proposal in spintronics is the Datta-Das `spin transistor' \cite{datta1990electronic}.  In the Datta-Das transistor, a spin-polarized current would be modulated through the Rashba spin-orbit interaction of electron carriers.  Although effects very similar to those proposed in the original paper by Datta and Das have finally been demonstrated \cite{koo2009control}, there is still much to be understood about, e.g., the role of Coulomb interactions in these devices (the subject of Sec.~\ref{sec:CoulombandSOI}). Intense research efforts on quantum dots in III-V semiconductors have also led to remarkable progress in single-spin initialization, manipulation, and readout.  All the basic ingredients for quantum computation (the so-called ``DiVincenzo criteria" \cite{DiVincenzo2000}) have been demonstrated in isolated (separate) experiments for spins in quantum dots.  However, there remain many questions about the viability of combining these ``nuts-and-bolts" in the presence of a random solid-state environment. These questions will be partially addressed in Sec.~\ref{sec:hole-echo-decay}.

In addition to nanoscale devices based on \emph{electron} spins, there is a growing interest in the spin properties of valence-band \emph{holes} in semiconductors.  We will focus our attention in this Chapter on holes in III-V semiconductors (GaAs, InAs, InSb, etc.), which share a common band structure. Much of the interest in hole-spin systems originates from the inherently strong spin-orbit coupling of the valence band.  This strong spin-orbit coupling introduces unique features for hole carriers arising from the orbital moment of the $p$-like valence band, relative to the $s$-like conduction band for electrons in these materials.  This review will therefore illustrate the specific advantages and disadvantages of hole-spin systems over their electron counterparts.  We will analyze these differences as they apply to both spin transport (Sec.~\ref{sec:CoulombandSOI}) as well as single-spin manipulation and decoherence (Sec.~\ref{sec:hole-echo-decay}).

In Sec.~\ref{non_interacting_hamiltonians} we give a brief review of the dominant spin-orbit terms for electrons and holes in two dimensions. Since local electrical gating is typically much easier to achieve experimentally than local magnetic fields, spin-orbit interactions represent an essential tool to address the spin degree of freedom at the nanoscale and a vast body of literature exists on the subject \cite{winkler2003spin,fabian2007semiconductor}. Most studies related to spin-orbit coupling are based on single-particle models, neglecting (or ignoring) the effects of Coulomb interaction. On the other hand, the role of electron-electron interactions in electron liquids is a well-developed and active field of research in itself \cite{giuliani2005quantum}, although most studies in this field neglect or ignore effects of spin-orbit coupling. A natural question that arises is then how the properties of such two-dimensional systems with strong spin-orbit interactions are modified by the presence of Coulomb repulsion, the subject of Sec.~\ref{sec:CoulombandSOI}. The need for a better understanding of this problem is highlighted by recent experiments on dilute hole liquids, in which strong correlations and spin-orbit coupling effects coexist. In this regime, these hole liquids show a significant deviation from the conventional behavior of two-dimensional electron liquids \cite{winkler2005anomalous, chiu2011effective}. 

Interaction effects can be computed accurately in the high-density limit by controlled perturbative methods (and some non-perturbative arguments have been put forward as well \cite{chesi2007two,aasen2012quasiparticle}).  In contrast, a quantitative understanding of the more correlated low-density regime including spin-orbit coupling is not yet possible. Various aspects of this problem have been examined in the theoretical literature, including quasiparticle properties (effective mass and lifetime) \cite{chen1999exchange,saraga2005fermi,nechaev2009hole,nechaev2010inelastic,chesi2011two,agarwal2011plasmon,aasen2012quasiparticle}, screening and plasmon excitations \cite{wang2005plasmon,pletyukhov2006screening,badalyan2009anisotropic,agarwal2011plasmon}, and more exotic collective excitations, including chiral spin waves \cite{ashrafi2012chiral}. Other works have considered ground-state properties including the repopulation of the spin bands \cite{chesi2007exchange}, the Hartree-Fock theory of broken-symmetry spin-polarized states \cite{giuliani2005on,chesi2007effects,juri2008hartree}, and low-density inhomogeneous phases \cite{berg2012electronic}. Studies of the ground-state energy including correlation effects can be found in Refs.~\cite{ambrosetti2009quantum,chesi2011two,chesi2011high-density}, while effects of interactions on the spin susceptibility with Rashba spin-orbit coupling have been studied in Refs.~\cite{dimitrova2005spin,chesi2007effects,agarwal2011plasmon,zak2010spin,zak2011ferromagnetic}, where several types of non-analytic correction have also been computed \cite{zak2010spin,zak2011ferromagnetic}. We would also like to mention the recent surge of interest on topological insulators and schemes supporting localized Majorana modes \cite{hasan2010colloquium,qi2011topological,lutchyn2010majorana,oreg2010helical,alicea2010majorana,alicea2011non,Mourik2012,Williams2012,Rokhinson2012}. While these systems are characterized by strong spin-orbit couplings, a proper understanding of interaction effects is an emerging and interesting problem in these fields as well (see, e.g., \cite{Gangadharaiah2011,Stoudenmire2011interaction}, for Majorana modes).
 
Part of the above mentioned work (that concerned with hole physics) is reviewed in Sec.~\ref{sec:CoulombandSOI} of this Chapter. In Sec.~\ref{sec:CoulombandSOI}, we will pay particular attention to the behavior of the exchange-correlation energy and quasiparticle excitations by contrasting expected effects in electron and hole liquids. As we will discuss, the group velocities of quasiparticles at the Fermi energy have quite distinct features for hole systems in the presence of spin-orbit coupling. A particularly vivid demonstration of this difference is provided by a hole wavepacket moving ballistically in a two-dimensional system. As we discuss in Sec.~\ref{sec:CoulombandSOI}, an initially spin-unpolarized hole wavepacket at the Fermi energy will spatially separate into its two spin components \cite{aasen2012quasiparticle}. In contrast, electrons moving in the presence of conventional Rashba and Dresselhaus spin-orbit interactions will remain spin-unpolarized.


While spin transport and manipulation are important for potential spintronic devices, quantum coherence and control are necessary for potential implementations of spin-based quantum computing, as well as more general quantum information processing tasks.  Hole spins confined to semiconductor quantum dots have several potential advantages from the viewpoint of quantum coherence as well.  The main difference between hole spins and electron spins can be traced once again to strong spin-orbit coupling.  The result of this coupling is to lock the orbital angular momentum $l=1$ to the spin angular momentum, $s=1/2$.  The states at the top of the valence band (the heavy holes) then transform like states with angular momentum $J_z = \pm 3/2$~\cite{Bulaev2005}. A primary decoherence source for electron spins is the Fermi contact hyperfine interaction between an electron spin and surrounding nuclear spins~\cite{coish2004hyperfine,klauser2006nuclear,Stepanenko2006,Greilich2007,Reilly2008}.  For holes, the difference in the microscopic spin-orbital state drastically modifies its coupling to the nuclear spin bath, and consequently, its decoherence properties~\cite{fischer2008spin,Eble2009}.  Heavy-hole spins in quantum dots are only affected by a weaker and strongly anisotropic hyperfine interaction~\cite{fischer2008spin}, leading to longer coherence times in a transverse magnetic field~\cite{fischer2008spin} or for a ``narrowed" nuclear-spin bath state~\cite{coish2004hyperfine,klauser2006nuclear}.  

Although single-\emph{electron} spin coherence measurements are becoming more common, single-\emph{hole} spin coherence is a relatively new frontier with many as-yet unanswered questions.  Recent experiments have determined the value~\cite{Chek2011,Fallahi2010} and sign~\cite{Chek2013} of the hyperfine coupling constants for hole spins and have achieved coherent optical control of the hole spin in single~\cite{DeGreve2011,Godden2012prl} and double~\cite{Greilich2011} quantum dots, thus enabling measurements of the relaxation time $T_1$~\cite{Heiss2007,Gerardot2008, Dahbashi2012}, estimates of a bound on the ensemble-averaged free-induction decay time $T_2^*$~\cite{Brunner2009}, and more recently, real-time measurements of $T_2^*$ and the spin-echo decay time, $T_2$~\cite{DeGreve2011}.  In Sec.~\ref{sec:hole-echo-decay} we will focus on a review of recent theoretical work exploring related hole-spin-echo decay in experimentally accessible regimes~\cite{Wang2012}.

\section{Types of spin-orbit coupling}\label{non_interacting_hamiltonians}

Spin-orbit couplings in semiconductors are discussed in several excellent books and reviews~\cite{zutic2004spintronics,fabian2007semiconductor,awschalom2002semiconductor,winkler2003spin}. Here we will restrict ourselves to a brief outline of the main spin-orbit mechanisms at play in III-V quantum wells, by emphasizing an important difference between electron and hole systems: while for electrons the dominant spin-orbit coupling mechanisms are linear in momentum, the dominant spin-orbit terms for heavy holes have a non-linear (quadratic or cubic) dependence on momentum.  This dependence has been confirmed in the literature on heavy holes, both theoretically and experimentally. As we will see in Section~\ref{sec:CoulombandSOI}, this distinction is important in discussing the interplay of spin-orbit and Coulomb interactions in a two-dimensional liquid.  Quantities that are modified by this interplay include the exchange-correlation energy and the effective mass and lifetime of quasiparticle excitations.

\subsection{Rashba and Dresselhaus spin-orbit interactions}

For two-dimensional electrons (with motion confined to the $x-y$ plane) the two main contributions to the spin-orbit interaction are well known.  Adding these to the kinetic energy,
\begin{equation}\label{H_electrons}
H=\frac{p^2}{2m}+\alpha (p_y \sigma_x - p_x \sigma_y) + \beta (p_y \sigma_y -p_x \sigma_x),
\end{equation}
where $\mathbf{p}$ is the momentum operator, $m$ the band mass, and $\boldsymbol{\sigma}$ is the vector of Pauli matrices. The first spin-orbit term (proportional to $\alpha$) is the Rashba spin-orbit coupling~\cite{bychkov1984properties,bychkov1984oscillatory}. This term has the same structure as the Dirac spin-orbit coupling $\propto \vec{\nabla} V({\bf r}) \cdot (\boldsymbol{\sigma}\times {\bf p})$, if the potential $V({\bf r})$ describes an electric field $E_z \hat{z}$. In fact, this type of spin-orbit coupling is caused by an asymmetry in the confinement potential (along $z$) which defines the quantum well. To derive this contribution, one can apply perturbation theory to a multiband Hamiltonian for the envelope functions.  This is the usual procedure applied to the 8-band Kane model (see, for example~\cite{winkler2003spin,deAndrada1994}). This procedure yields an estimate of the coupling $\alpha$ as follows
\begin{equation}\label{alpha_formula}
\hbar \alpha =\frac13 \left(\frac{1}{E_0^2}- \frac{1}{(E_0+\Delta_0)^2}\right) P^2 e E_z,
\end{equation}
where we use the notation of Ref.~\cite{winkler2003spin} for the band structure parameters: $E_0$ is the main gap of the III-V semiconductor, $\Delta_0$ is the spin-orbit splitting at the top of the valence band, and $m_0P/\hbar$ is a matrix element of the momentum operator between the valence and conduction bands (with $m_0$ the free-electron mass). Note that $\alpha$ would be zero for $\Delta_0=0$. For III-V semiconductors, $P\sim 1$ eV-nm, which gives $\hbar \alpha \sim 0.05$ meV-nm for GaAs and $\hbar \alpha \sim$ 15 meV-nm for InAs, using a value of the electric field $E_z \simeq 1~{\rm V}/\mu$m.

The second spin-orbit term in Eq.~(\ref{H_electrons}) (that proportional to $\beta$) is the Dresselhaus spin-orbit coupling~\cite{dresselhaus1955spin,dyakonov1986spin}. This term arises from the bulk asymmetry of III-V semiconductors and for this reason it is already present in the three-dimensional band structure with the following form~\cite{dresselhaus1955spin}:
\begin{equation}\label{dresselhaus_3D}
H_{D}=\gamma_{D} [ (p_y^2 -p_z^2)p_x \sigma_x+(p_z^2 -p_x^2)p_y \sigma_y+(p_x^2 -p_y^2)p_z \sigma_z],
\end{equation}
where $x,y,z$ are along the crystallographic axes.
An estimate of $\gamma_D$ can be obtained by applying perturbation theory to an extended Kane model (including 14 bands), which leads to a formula analogous to Eq.~(\ref{alpha_formula}). This procedure yields $\hbar^3 \gamma_D \simeq 27$ meV-$\mathrm{nm}^3$ for both GaAs and InAs~\cite{winkler2003spin}. Although Eq.~(\ref{dresselhaus_3D}) has been known for a very long time, there is some recent controversy related to the accuracy of this value for $\gamma_D$ in GaAs. A significantly smaller value $\hbar^3 \gamma_D  \simeq 9$ meV-$\mathrm{nm}^3$ was obtained in Ref.~\cite{krich2007cubic}, from a theoretical analysis of conductance measurements through open quantum dots~\cite{zumbuhl2002spin,zumbuhl2005conductance}. A collection of theoretical and experimental values for $\hbar^3 \gamma_D$ can also be found in Ref.~\cite{krich2007cubic}. On the other hand, good agreement with spin-relaxation measurements in the impurity band of GaAs has been obtained in Ref.~\cite{intronati2012spin}, using $\hbar^3 \gamma_D  \simeq 27$ meV-$\mathrm{nm}^3$. 

Starting from Eq.~(\ref{dresselhaus_3D}), we can see that the two-dimensional Dresselhaus coupling in Eq.~(\ref{H_electrons}) can be derived by substituting $p_z,p_z^2$ by the expectation values $\langle p_z \rangle=0 $ and $\langle p_z^2 \rangle \simeq (\pi \hbar/W)^2$, where $W$ is the width of the quantum well, and neglecting small terms that are cubic in the in-plane momentum. Using $\hbar^3 \gamma_D  = 27$ meV-$\mathrm{nm^3}$ and $W=20$ nm gives $\hbar \beta \simeq \hbar^3\gamma_D (\pi/W)^2 \simeq 0.7$ meV-$\mathrm{nm^3}$. So, with these specific values of $E_z$ and $W$, we find that for GaAs the Dresselhaus spin-orbit coupling is dominant over the Rashba mechanism, while for InAs the opposite is true. This derivation also emphasizes that Eq.~(\ref{H_electrons}) only applies if the growth direction of the quantum well is $[001]$, i.e., along a crystallographic direction~\cite{dyakonov1986spin}. From Eq.~(\ref{dresselhaus_3D}) we can also derive the analogue of Eq.~(\ref{H_electrons}) for different growth direction of the quantum well. An interesting case (see, e.g., Ref.~\cite{sih2005spatial}) is the growth direction $[110]$, for which the Dresselhaus coupling becomes simply proportional to $\sigma_x + \sigma_y$, thus the spin is conserved along the quantum well growth direction.

Several new phenomena arise from the simultaneous presence of both Rashba and Dresselhaus couplings.  These include an anisotropic conductivity~\cite{chalaev2008anisotropic}, the formation of `electron-beams'~\cite{berman2010electron}, and a beating of Friedel oscillations~\cite{badalyan2010beating}, to mention just a few examples. In general, the presence of both Rashba and Dresselhaus terms results in a Hamiltonian which is anisotropic in the $x-y$ plane. The special (symmetric) case $\alpha=\beta$ results in a new conservation law~\cite{schliemann2003nonballistic}.  One consequence of this conservation law is the existence of long-lived spin textures (the so-called `spin helix')~\cite{bernevig2006exact,koralek2009emergence,Walser2012direct}. Determining the precise values of $\alpha$ and $\beta$ in any given sample is often challenging.  The ratio $\alpha/\beta$ can be measured through the spin-galvanic effect~\cite{ganichev2002spin,ganichev2004experimental}, while it is also possible to extract the individual values of $\alpha$ and $\beta$ independently using Faraday rotation to monitor the precession of electron spins as they move through a sample~\cite{meier2007measurement}. 

A large body of literature considers the effect of pure Rashba spin-orbit interaction ($\beta=0$), which is isotropic in the $x-y$ plane. To justify this assumption, one can notice that the pure Rashba Hamiltonian is equivalent to the case of pure Dresselhaus ($\alpha=0$), after performing a suitable rotation in spin space $\sigma_{x,y}\to \sigma_{y,x}$. Thus, it is sufficient that one of the two spin-orbit mechanisms is dominant over the other. This assumption is often reasonable since, as we have seen, the relative strength of $\alpha$ and $\beta$ depends on the specific choice of materials and the electronic confinement along $z$. For example, the Rashba spin-orbit coupling is absent in symmetric quantum wells. Furthermore, other material systems exist where the Rashba spin-orbit coupling is important and the Dresselhaus term might vanish. The presence of a linear-in-momentum spin splitting has been revealed in angle-resolved photoemission spectra from the Shockley surface state on Au(111)~\cite{lashell1996spin}. In that case, spin-resolved photoemission experiments~\cite{hoesch2004spin} directly reveal the spin structure of the electronic states, which are found to agree well with the Rashba eigenstates [see Eq.~(\ref{eq:Ek-non-interacting}), below]. Furthermore, in the quasi-two-dimensional electron gas at the ${\rm LaAlO}_3\rm{/SrTiO}_3$ interface~\cite{ohtomo2004high}, a tunable Rashba spin-orbit interaction can coexist with superconductivity~\cite{caviglia2008electric,caviglia2010tunable}. For this system, the existence of persistent spin oscillations was predicted in Ref.~\cite{agarwal2011persistent}. The presence of Rashba spin-orbit coupling also plays an important role in thin asymmetric ferromagnetic layers in contact with heavy elements, including Pt/Co/Al${\rm O}_x$ structures~\cite{miron2010current,pi2010tilting,miron2011perpendicular,miron2011fast}.

\subsection{Generalized spin-orbit coupling}\label{non-linear spin-orbit}

We can generalize the discussion of the last section by considering the following model Hamiltonian~\cite{chesi2007exchange,aasen2012quasiparticle}:
\begin{equation}\label{eq:H0}
H_{0} = \frac{p^2}{2m} + i\gamma \frac{p_{-}^n\sigma_{+} -p_{+}^n\sigma_{-} }{2},
\end{equation}
where $p_{\pm} = p_{x} \pm i p_{y}$, $\sigma_{\pm} = \sigma_{x} \pm i \sigma_{y}$. Introducing Eq.~(\ref{eq:H0}) is useful since for $n=1,2,3$ it represents different realizable spin-orbit coupling terms. In particular, the $n=1$ case of Eq.~(\ref{eq:H0}) is equivalent to the Rashba term [$\propto \alpha$ in Eq.~(\ref{H_electrons})]. Alternatively, the $n=2,3$ Hamiltonians are appropriate, for example, for heavy holes in III-V semiconductor quantum wells. With $n=3$, Eq.~(\ref{eq:H0}) has been shown to describe the main contribution to the spin-orbit coupling for holes in the presence of an asymmetric confinement potential~\cite{winkler2000rashba}, in good agreement with magnetoresistance measurements~\cite{winkler2002anomalous}. The dominant $\sim p^3$ dependence can be related to the total angular momentum of the heavy holes. While electrons in the $s$-like conduction band have angular momentum $\pm 1/2$ (from the spin), the $\uparrow/\downarrow$ states of the effective heavy-hole Hamiltonian (\ref{eq:H0}) with $n=3$ correspond to $J_z=\pm 3/2$.  This is a consequence of the $p$-like character of the valence band, since the orbital angular momentum combines with the spin into $J=3/2$ states and the confinement along $z$ further splits the degeneracy between heavy holes ($J_z=\pm 3/2$) and light holes ($J_z=\pm 1/2$). Furthermore, the coupling for $n=2$ is present for heavy-hole systems with a finite in-plane magnetic field $B$ and $\gamma \propto B$~\cite{winkler2003spin,bulaev2007electric,chesi2007exchange}, consistent with considerations on time-reversal symmetry (i.e., the Hamiltonian must be invariant under the simultaneous change $\mathbf{p}\to -\mathbf{p}$, $\boldsymbol{\sigma}\to -\boldsymbol{\sigma}$, $B\to-B$). While other types of spin-orbit coupling mechanisms exist for holes (see, e.g., Ref. \cite{rashba1988spin} for a discussion of symmetric quantum wells), Eq.~(\ref{eq:H0}) represents a useful model which often captures the dominant contributions. For example, recently an anomalous behavior of polarization in magnetic focusing experiments for holes~\cite{rokhinson2004spin,chesi2011anomalous} has been successfully interpreted on the basis of the $n=3$ Hamiltonian of Eq.~(\ref{eq:H0}). 
\begin{figure}[t]
  \centering
 \includegraphics[width = 0.9\textwidth]{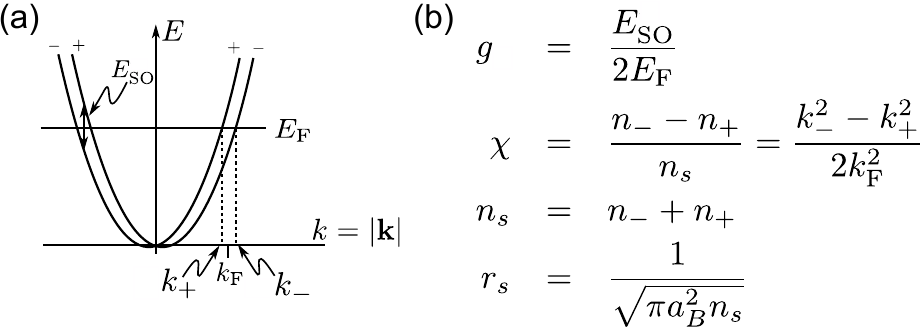}\\
   \caption{\label{fig:ChiralBands} Chiral energy bands of a spin-orbit coupled system and parameters characterizing the strength of the spin-orbit coupling, $g$, the Coulomb interaction, $r_s$, and the chirality, $\chi$.  These parameters are written in terms of the densities, $n_\pm$, and Fermi wavevectors $k_\pm$ of the two chiral bands, and the effective Bohr radius, $a_B$.}
\end{figure}

Diagonalizing $H_0$ yields the following eigenstates, with the corresponding energies: 
\begin{equation}\label{eq:Ek-non-interacting}
\varphi_{ \mathbf{k} \pm}(\mathbf{r}) = \frac{ e^{i \mathbf{k} \cdot \mathbf{r} }}{\sqrt{2 L^2}} \begin{pmatrix} 1 \\ \pm i e^{in \theta_{\mathbf{k}}} \end{pmatrix}, \quad 
E_{\pm}^{0} (k) = \frac{\hbar^2 k^2}{2m} \pm \gamma \hbar^n k^n,
\end{equation}
where $\mathbf{k}$ is a wavevector in the $x-y$ plane, $\theta_{\bf k}$ is the angle $\mathbf{k}$ makes with the $x$-axis, $L$ is the linear size of the system, and $\pm$ labels the two chiral spin branches. We will generally assume $\gamma \geq 0$, such that the $``-"$ subband has lower energy.


In the absence of spin splitting, the Fermi wavevector is $k_F =\sqrt{2 \pi n_s}$, with $n_s$ the total density. However, for $\gamma\neq 0$, the densities $n_\pm$ of the two non-degenerate spin subbands are different (with $n_s=n_++n_-$). In this general case, it is useful to write the two distinct Fermi wavevectors $k_\pm$ (see Fig.~\ref{fig:ChiralBands}) as follows:
\begin{equation}\label{eq:kpm}
k_\pm = k_F \sqrt{|1 \mp \chi |}.
\end{equation}
When $0\leq \chi< 1$, there are two Fermi circles with radii $k_\pm$ and we have that $\chi=(n_- - n_+)/n_s$. Thus, $\chi$ coincides in this case with the chirality of the system.  The parameter $\chi$ is analogous to the fractional polarization of spin-polarized systems (although in this case the spin quantization axis depends on the direction of $\mathbf{k}$ and so there is no definite spin polarization when all states of the same $|\mathbf{k}|$ are filled uniformly for all directions in $k$-space). 

To describe the non-interacting properties, and more specifically the value of $\chi$, we note that the Hamiltonian (\ref{eq:H0}) can be characterized for all $n=1,2,3$ by the ratio of the spin-orbit energy splitting $E_{SO}=2\gamma\hbar^n k_F^n$ to the Fermi energy $E_F=\hbar^2 k_F^2/2m$. We thus define the dimensionless parameter:
\begin{equation}
g=\frac{\gamma \hbar^n k_F^n}{E_F},
\end{equation}
which allows us to write the ground-state occupancy as
\begin{equation}\label{pmin123}
\chi_0(g) =\left\{
\begin{array}{cl}
g \sqrt{1-\frac{g^2}{4}},   & ~\qquad n=1, \\
g  ,   & ~\qquad n=2, \\
g \sqrt{
\frac{-3 g^4+6 g^2 - 2+2
(1-2 g^2)^{3/2}}{g^6}}.  & ~\qquad n=3,
\end{array}
\right.
\end{equation}
In all three cases, the leading dependence at small $g$ is simply given by $\chi_0 \simeq g$.

In writing Eq.~(\ref{pmin123}), we have assumed the most common situation of a rather small $g$, such that $\chi_0(g)<1$. The limit of large $g$ requires a separate discussion for $n=1$, due to the fact that $E_-^0(k)$ of Eq.~(\ref{eq:Ek-non-interacting}) has its minimum at a finite wavevector, $k=m\gamma/\hbar$, rather than at $k=0$. Thus, for $n=1$ and sufficiently low density, the occupation is in the $``-"$ subband and forms a ring in momentum space. In this case, Eq.~(\ref{eq:kpm}) maintains its validity but $\chi\geq 1$ and $k_\pm$ now determine the inner and outer radii ($k_-<m\gamma/\hbar<k_+$) for the ring of occupied states. The precise condition for this regime is $g \geq \sqrt{2}$, for which we can write $\chi_{0}(g)=g^2/4+1/g^2\geq 1$. Note that, since the chirality for a single occupied band is simply $(n_- - n_+)/n_s=1$, the formula $\chi=(n_- - n_+)/n_s$ in Fig.~\ref{fig:ChiralBands} is no longer valid when $\chi>1$. It is then more appropriate to refer to $\chi$ as a `generalized chirality'. On the other hand, both spin subbands $E_\pm^0(k)$ of Eq.~(\ref{eq:Ek-non-interacting}) have minimum energy at $k=0$ if the spin-orbit splitting is quadratic or cubic in momentum (for $n=2,3$, respectively). For $n=2,3$, and in the limit of small densities, the occupation of the two chiral bands is therefore always determined by two disks with radii given by $k_\pm$ in Eq.~(\ref{eq:kpm}).

\section{Spin-orbit coupling and Coulomb interactions}\label{sec:CoulombandSOI}

We now turn to many-body effects and discuss a few significant consequences of the interplay between spin-orbit coupling and Coulomb interaction. For the same non-interacting problem discussed in Sec.~\ref{non-linear spin-orbit}, we consider here the following Hamiltonian  ($n=1,2,3$),
\begin{equation}\label{eq:H}
H=\sum_i \left[ \frac{p_i^2}{2m}+ i\gamma \frac{p_{i,-}^n\sigma_{i,+} -p_{i,+}^n\sigma_{i,-} }{2} \right] +\frac12 \sum_{i\neq j} \frac{e^2}{\epsilon_r |{\bf r}_i -{\bf r}_j|},
\end{equation}
where $\epsilon_r$ is the dielectric constant and standard terms associated with the neutralizing background have been omitted for simplicity~\cite{giuliani2005quantum}. The presence of the Coulomb repulsion introduces another dimensionless parameter in the problem in addition to $g$:
\begin{equation}
r_s =\frac{1}{\sqrt{\pi a_B^2 n_s}},
\end{equation}
where $a_B=\epsilon_r \hbar^2/m e^2$ is the effective Bohr radius. The Wigner-Seitz radius $r_s$ characterizes the strength of the Coulomb interaction with respect to $E_F$: in the small-$r_s$ limit (approached at large density $n_s$), the Coulomb interaction can be treated perturbatively, while at large $r_s$, strong correlation effects are present~\cite{giuliani2005quantum}. The most studied example of Eq.~(\ref{eq:H}) is certainly the electron liquid with pure Rashba spin-orbit coupling ($n=1$). However, this is also a special case in which several many-body effects are left essentially unaffected by the presence of spin-orbit coupling. As discussed below, this behavior can be contrasted to hole systems with a dominant spin-orbit coupling that is quadratic or cubic in momentum ($n=2,3$), for which special cancellations that are present for $n=1$ do not occur.

\subsection{Exchange-correlation energy}\label{noeffect_couplimb_rashba}

One of the most fundamental properties of the Hamiltonian, Eq.~(\ref{eq:H}), is the ground-state energy. By considering the many-body ground state with Fermi surfaces determined by $\chi$ as in Eq.~(\ref{eq:kpm}), we can define the corresponding  exchange-correlation energy $\mathcal{E}_{xc}(g,r_s,\chi)$ as follows~\cite{chesi2011high-density} (in effective Rydbergs),
\begin{equation}\label{total_energy}
\mathcal{E}(g , r_s, \chi)=\frac{1+\chi^2}{r_s^2}- \frac{2g}{r_s^2}\frac{\sqrt{|1+\chi|^{n+2}}-\sqrt{|1-\chi|^{n+2}}}{n+2}+\mathcal{E}_{xc}(g,r_s,\chi).
\end{equation}
Here, the first two terms are, respectively, the non-interacting kinetic and spin-orbit energies. Note that Eq.~(\ref{total_energy}) is written at fixed $\chi$ and to find the ground-state energy it is necessary to further minimize $\mathcal{E}(g , r_s, \chi)$ with respect to $\chi$. For example, $\chi_0(g)$ of Eq.~(\ref{pmin123}) is obtained by minimizing the non-interacting part of Eq.~(\ref{total_energy}), i.e., that without $\mathcal{E}_{xc}$~\cite{chesi2007exchange}.\footnote{For $n=1$ and $\chi\geq 1$ one should replace $1+\chi^2$ with $2\chi$ in the non-interacting kinetic energy.}

As mentioned above, the $n=1$ case is the most-commonly studied case. A general result for $\mathcal{E}(g,r_s,\chi)$ in this case has been obtained from an analysis of the perturbative expansion of $\mathcal{E}_{xc}(g,r_s,\chi)$, to all orders in the Coulomb interaction parameter $r_s$. By Taylor expanding a generic energy diagram in $g,\chi$, it has been found that the quadratic term is always $\propto(g-\chi)^2$~\cite{chesi2011two}. Performing a similar expansion on the non-interacting energy of Eq.~(\ref{total_energy}) gives
\begin{equation}\label{exact_gchi_dependence}
\mathcal{E}(g,r_s,\chi)= \frac{1}{r_s^2}+ \mathcal{E}_{xc}(r_s)-\frac{g^2}{r_s^2}+\left(\frac{1}{r_s^2}+ C\right) (g-\chi)^2 + \ldots,
\end{equation}
with $C$ an unspecified constant which depends only on $r_s$, while the exchange-correlation energy without spin-orbit coupling, $\mathcal{E}_{xc}(r_s)$, is a well-studied quantity~\cite{rajagopa1977correlations,giuliani2005quantum,attaccalite2002correlation,chesi2007correlation,drummond2009quantum,loos2011exact}. Provided the constant satisfies $C>-1/r_s^{2}$ (which is always true at sufficiently high density), Eq.~(\ref{exact_gchi_dependence}) shows that the total energy is minimized by the same value of the chirality $\chi_0 \simeq g$ as for the non-interacting system, see Eq.~(\ref{pmin123}). An immediate consequence of setting $\chi \simeq g$ is that the quadratic corrections to the ground-state energy are generally absent for $n=1$. 

An explicit evaluation of $\mathcal{E}_{xc} (r_s, g,\chi)$ becomes possible in the high-density limit, in terms of a controlled expansion in powers of the small parameter $r_s$. In that regime, the leading contribution is given by the exchange energy~\cite{chesi2011high-density}:
\begin{equation}\label{exchange_energy_n=1}
\mathcal{E}_{xc} (r_s, g,\chi) \simeq -\frac{8\sqrt{2}}{3\pi r_s}+\frac{\mathcal{E}_x (\chi)}{r_s} = -\frac{8\sqrt{2}}{3\pi r_s}+\frac{\sqrt{2}}{48 \pi r_s}\chi^4 \left(\ln \frac{\chi}{8}+\frac{23}{12} \right) + \ldots,
\end{equation}
which is in agreement with Eq.~(\ref{exact_gchi_dependence}), for $C=0$. Higher-order correlations have also been examined, confirming the general form of Eq.~(\ref{exact_gchi_dependence}) and that the coefficient $C$ is generally non-zero. The second-order correlation energy yields a density-independent prefactor which is $C=0$ for the exchange contribution, but $C=-1/4$ for the direct term ~\cite{chesi2011high-density}. The next contribution to the correlation energy and the corresponding expression for $C$, from the infinite resummation of ring diagrams, have also been studied explicitly~\cite{chesi2011high-density}. However, these higher-order terms do not modify the leading result for the ground-state energy, obtained from Eq.~(\ref{exchange_energy_n=1}) by setting $\chi=g$. The possibility of non-analytic corrections to the energy $\propto g^3 \ln^2 g$, which are still compatible with Eq.~(\ref{exact_gchi_dependence}), has been discussed in Ref.~\cite{zak2011ferromagnetic}. 

\begin{figure}[t]
  \centering
\includegraphics[width = 0.9\textwidth]{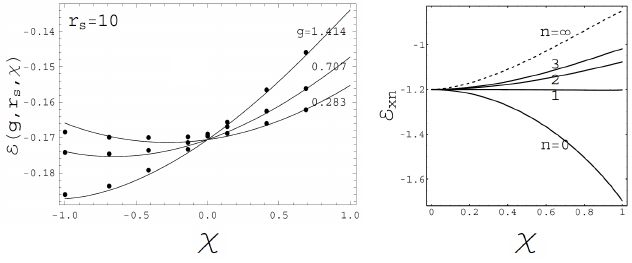}
   \caption{\label{fig:QMCdata} Left (from Ref.~\cite{chesi2011high-density}): Quantum Monte Carlo energies (dots) for the electron liquid at $r_s=10$ including the Rashba spin-orbit couplings~\cite{ambrosetti2009quantum}. The solid lines are obtained using the exchange-correlation energy in the absence of spin-orbit coupling. Right (from Ref.~\cite{chesi2007exchange}): exchange energy at $r_s=1$ for eigenstates of the non-interacting Hamiltonian (\ref{eq:H0}), parameterized by the chirality $\chi$.}
\end{figure} 

A striking confirmation of the small effect of spin-orbit coupling extending to the non-perturbative regime comes from an analysis of the exchange-correlation energy obtained by making use of the diffusion Quantum Monte Carlo method~\cite{ambrosetti2009quantum}. Fig.~\ref{fig:QMCdata} reproduces some of the numerical data (dots), for representative values of the parameters $g,r_s,\chi$, compared with the simple approximation (solid lines) of setting $\mathcal{E}_{xc}(g,r_s,\chi) \simeq \mathcal{E}_{xc}(r_s)$ in Eq.~(\ref{total_energy})~\cite{chesi2011high-density}. As seen in Fig.~\ref{fig:QMCdata}, this approximation is in excellent agreement with the Quantum Monte Carlo results, even in the strongly correlated regime $r_s > 1$ and at large values of the spin-orbit coupling parameters $g,\chi \sim 1$.  

These arguments thus indicate that the exchange-correlation energy is always essentially unchanged at finite $g$ for $n=1$. On the other hand, this behavior is peculiar to the paramagnetic state with $n=1$ Rashba spin-orbit coupling.  In contrast, a significant interplay between Coulomb interaction and Rashba spin-orbit coupling exists for broken-symmetry states~\cite{giuliani2005on,chesi2007effects,juri2008hartree,berg2012electronic} and in the presence of a magnetic field (e.g., for the spin susceptibility~\cite{zak2010spin,agarwal2011plasmon,zak2011ferromagnetic}). Furthermore, non-linear-in-momentum spin-orbit coupling terms have been studied where larger modifications of the exchange-correlation energy can be found. In particular, the exchange energy has been calculated in Ref.~\cite{chesi2007exchange} and is plotted in Fig.~\ref{fig:QMCdata} (at $r_s=1$). In contrast to the $n=1$ result, the $n=2,3$ curves have a much stronger dependence on $\chi$. In fact, a small-$\chi$ expansion gives
\begin{equation}\label{exchange_energy_n=123}
\mathcal{E}_{x} (r_s, g) \simeq -\frac{8\sqrt{2}}{3\pi r_s}+\frac{\sqrt{2}}{\pi r_s}\left(\sum_{m=0}^n \frac{1}{2m-1} \right)\chi^2  + \ldots.
\end{equation}
While the $\sim \chi^2$ correction vanishes for $n=1$, in agreement with Eq.~(\ref{exchange_energy_n=1}), a finite result is obtained with $n=2,3$.

Furthermore, we can compare the $n=2,3$ exchange energies to the usual result for a spin-polarized system, which is plotted in Fig.~\ref{fig:QMCdata} as the $n=0$ curve\footnote{For $n=0$, the parameter $\chi$ has the meaning of a fractional spin polarization.}. As is known, the exchange energy favors spin-polarized states and the $n=0$ curve is thus a decreasing function of spin polarization. On the other hand, the exchange energy with $n=2,3$ has the opposite effect of \emph{suppressing} the subband population difference. In a realistic situation, both spin-orbit coupling and the Zeeman term coexist. On this basis, it was proposed in Ref.~\cite{chesi2007exchange} that a competition between the opposing effects of spin polarization and spin-orbit couping (induced by the magnetic field itself) could be responsible for the apparent suppression of interaction effects on the subband populations, observed for dilute hole systems in strong magnetic fields~\cite{winkler2005anomalous}.

\subsection{Spin-dependent quasiparticle velocity}\label{interacting holes}

Another remarkable effect of electron interactions is the renormalization of quasiparticle properties.  The combined effects of Coulomb interactions and  linear-in-momentum (Rashba and/or Dresselhaus) spin-orbit coupling have been studied in Refs.~\cite{chen1999exchange,saraga2005fermi,nechaev2009hole,nechaev2010inelastic,agarwal2011plasmon,aasen2012quasiparticle} with a variety of approximation schemes. In particular, perturbative expressions for the lifetime, effective mass, and spectral weight $Z$, have been obtained in Ref.~\cite{saraga2005fermi} within the RPA approximation. As in our previous discussion of the exchange-correlation energy, it was found in Ref.~\cite{saraga2005fermi} that all leading corrections (linear in $g$) are absent. This result is valid more generally since a non-perturbative argument analogous to that leading to Eq.~(\ref{exact_gchi_dependence}) holds for quasiparticles as well~\cite{chesi2011two,aasen2012quasiparticle}. 

\begin{figure}[t]
  \centering
\includegraphics[width = 0.45\textwidth]{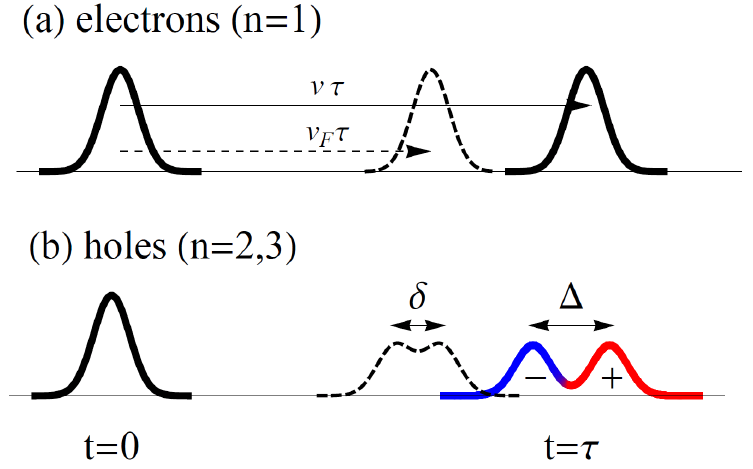}\hspace{0.5cm}
\raisebox{0.3cm}{\includegraphics[width = 0.45\textwidth]{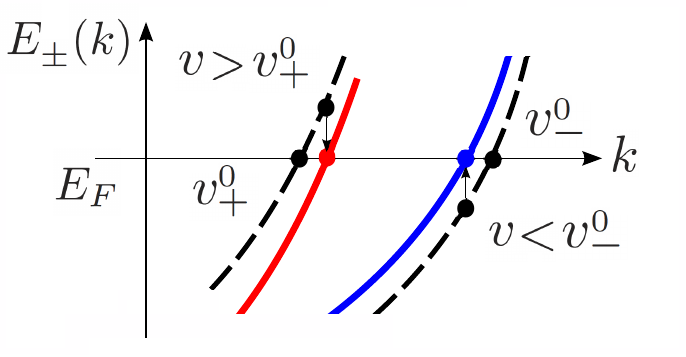}}
   \caption{\label{fig:ElectronsVsHoles}  Left (adapted from Ref.~\cite{aasen2012quasiparticle}): Schematic representation of the motion of an initially unpolarized wavepacket for a fixed time $\tau$. (a) refers to $n=1$ and (b) to $n=2,3$. At high density, the Coulomb interaction renormalizes the average velocity as well as the velocity difference between spin components, with respect to the non-interacting values (dashed wavepackets). Right: Schematic of the effect of repopulation on the spin-dependent velocities. The Coulomb interaction induces energy shifts of the non-interacting spin branches (dashed) which both reduce the chirality $\chi$ and contribute to the increase (decrease) of the non-interacting velocity $v_+^0~(v_-^0)$~\cite{aasen2012quasiparticle}. In addition to the energy shift, the slope of the non-interacting energy dispersion is modified as well~\cite{janak1969g,aasen2012quasiparticle}, but this effect is not shown here. }
\end{figure}

In contrast to the linear-in-momentum ($n=1$) case, there is no fortuitous cancellation when $n=2,3$, as appropriate for hole systems.  This observation motivates a study of quasiparticles with $n=2,3$~\cite{aasen2012quasiparticle}. We focus here on quasiparticle propagation~\cite{aasen2012quasiparticle}. By considering the group velocity at the Fermi energy, the difference between linear and non-linear-in-momentum spin-orbit couplings becomes apparent even for a non-interacting system. At small $r_s,g$, we have
\begin{equation}\label{vpm}
\frac{v_\pm^0}{v_F} \simeq 1 \pm \frac{g}{2}(n-1),
\end{equation}
with $v_{F}=\hbar k_F/m$. Equation~(\ref{vpm}) implies that the motion of a spin-unpolarized wavepacket at the Fermi energy will be qualitatively different for the two cases ($n=1$ and $n=2,3$), as schematically illustrated in Fig.~\ref{fig:ElectronsVsHoles}. If $n=1$ we have $v_+^0 = v_-^0$ and the wavepacket will remain spin-unpolarized for all time.  In contrast, if $n=2,3$, the two spin components have different velocities and will spatially separate. 

As is well-known~\cite{giuliani2005quantum}, many-body effects affect the quasiparticle dispersion, modifying Eq.~(\ref{vpm}) for an interacting system. We thus introduce the following general notation~\cite{aasen2012quasiparticle}
\begin{equation}\label{vpm_general}
\frac{v_\pm}{v_F} \simeq 1 + \delta v(r_s) + \delta v_\pm^0(g) + \delta v_\pm (r_s,g),
\end{equation}
where $\delta v(r_s)$ is the known spin-independent contribution from the Coulomb interaction at $g=0$, which has been studied extensively in the literature~\cite{janak1969g,tan2005measurements,giuliani2005quantum,drummond2009quantum}. The non-interacting effect of the spin-orbit coupling is $\delta v_\pm ^0(g) \simeq \pm (n-1)g/2$, see Eq.~(\ref{vpm}), and the last term describes the nontrivial interplay between spin-orbit coupling and Coulomb interactions. The term $\delta v_\pm (r_s,g)$ has been studied in Ref.~\cite{aasen2012quasiparticle} within the context of the screened Hartree-Fock approximation~\cite{janak1969g,chen1999exchange} which gives, for $n=1$,
\begin{equation}\label{dvpm_n1}
\delta v_\pm (r_s,g) \simeq \frac{\sqrt{2}r_s}{16 \pi}\left[ g^2 \left(\ln \frac{g}{8}+ \frac32 \right)\pm \frac{g^3}{6} \left(\ln \frac{g^7}{8^5}+ \frac{319}{20} \right)+\ldots \right].
\end{equation}
The result of Eq.~(\ref{dvpm_n1}) is computed to lowest order in $r_s$, as an expansion in powers of small $g$. As seen in that equation, the leading term of $\delta v_\pm (r_s,g)$ contains a factor $g^2$ and is independent of the spin branch. The result is in agreement with the RPA analysis of Ref.~\cite{saraga2005fermi} [only the $O(g ^2\ln g)$ term was obtained there].  This agreement supports using the much simpler screened Hartree-Fock theory in the high-density regime. In Eq.~(\ref{dvpm_n1}), the $\sim g^3$ term introduces a velocity difference between the two subbands, which is a qualitatively new effect introduced by the Coulomb interaction~\cite{aasen2012quasiparticle}. In fact, $v_+^0 = v_-^0$ exactly for the non-interacting system. However, this finite velocity difference only arises as a subleading correction. As mentioned previously, the presence of non-linear-in-momentum spin-orbit coupling terms ($n=2,3$) induce much more dramatic effects than suggested by Eq.~(\ref{dvpm_n1}). Here we cite the corresponding result for $n=3$ to leading order in $r_s, g$~\cite{aasen2012quasiparticle}:
\begin{equation}\label{dvpm_n3}
\delta v_\pm (r_s,g) \simeq \pm \frac{2\sqrt{2}}{5 \pi} r_s g,~\qquad n=3.
\end{equation}
A similar expression is found for $n=2$ and in both cases $\delta v_\pm (r_s,g)\sim g$ and has opposite sign in the two chiral subbands. 

To better understand the velocity renormalization, here we consider the specific contribution to $\delta v_\pm (r_s,g)$ related to the repopulation effect mentioned after Eq.~(\ref{exchange_energy_n=123}). Minimization of the total energy, Eq.~(\ref{total_energy}), with respect to $\chi$ gives~\cite{chesi2007exchange}:
\begin{equation}\label{chi_renorm}
\chi \simeq g \left(1-\frac{\sqrt{2}r_s}{\pi }\sum_{m=0}^n \frac{1}{2m-1} \right)
\end{equation}
at small $g$ and $r_s$, when $\mathcal{E}_{xc}$ can be approximated as in Eq.~(\ref{exchange_energy_n=123}). While the effect of interactions is absent for $n=1$ (when we have $\chi \simeq g$ as in the non-interacting case), the factor in parenthesis reduces the value of $\chi$ for $n=2,3$. The same result, Eq.~(\ref{chi_renorm}), is obtained from the quasiparticle analysis of Ref.~\cite{aasen2012quasiparticle}, and is due to unequal energy shifts of the two spin branches, induced by the Coulomb interaction. As schematically illustrated in Fig.~\ref{fig:ElectronsVsHoles}, the reduced value of $k_-$ corresponds to a decrease of the non-interacting velocity in the $``-"$ (lower energy) subband, and the opposite is true for the $``+"$ subband. The sign of this effect is in agreement with Eq.~(\ref{dvpm_n3}). However, the final result also takes into account the wavevector dependence of the  interacting self-energy, which gives an additional spin-dependent contribution to $\delta v_\pm (r_s,g)$~\cite{aasen2012quasiparticle}. As it turns out, both effects of the Coulomb interaction have the same sign and contribute to \emph{enhance} the velocity difference. 

By referring to Fig.~\ref{fig:ElectronsVsHoles} and assuming $n=2,3$, we see how the non-interacting result is modified at high density, within the screened Hartree-Fock approximation. For a given travel time $\tau$, the total travel distance and the separation between the two spin components are larger than the non-interacting value ($\Delta > \delta$). The difference in Fermi velocity of the two spin branches could be experimentally probed with transport measurements of the effective mass~\cite{tan2005measurements,chiu2011effective} and Raman scattering experiments~\cite{jusserand1992zero}, as well as by direct optical imaging of the wave-packet separation. In fact, for a travel distance of order $\sim \mu$m, the travel time is $1-50$ ps (depending on density) and typical values of $\gamma$ for holes give spatial separations between the two spin components of a few hundred nm ($\sim 200$ nm for $g=0.1$ and $n=3$), which is within reach of Faraday rotation measurements~\cite{crooker1996terahertz,kikkawa1998resonant}. 

\section{Spin-echo decay in single-hole quantum dots} \label{sec:hole-echo-decay}

Spintronics research often involves studying the motion of spins in two-dimensional systems (the subject of the last section).  Carriers that are free to move in two dimensions can, however, have their spins efficiently randomized through momentum scattering in combination with spin-orbit interactions \cite{Meier1984optical}.  To achieve long spin coherence times, it is typically necessary to quench the orbital momentum by confining carriers to zero-dimensional quantum dots.  After freezing the orbital degrees of freedom, spin-orbit interactions may still limit hole-spin relaxation, but become an ineffective dephasing mechanism \cite{Bulaev2005,Trif2009}.  Although there are still situations where dephasing can be induced through spin-orbit coupling for quantum-dot-confined hole spins \cite{Lyanda-Geller2012}, these mechanisms can be suppressed through stronger two-dimensional confinement.

Even in the limit of extreme confinement, decoherence can still proceed through a small residual spin-spin coupling such as the hyperfine interaction between an electron or hole spin and surrounding nuclear spins in the lattice.  For a heavy-hole spin in a quasi two-dimensional quantum dot with negligible strain (see Ref.~\cite{tracy2013few} for very recent experimental progress), the hyperfine interaction $H_\mathrm{hf}$ with surrounding nuclei is approximately Ising-like~\cite{fischer2008spin,Fischer2010,Maier2012,Sinitsyn2012}:
\begin{equation}
\label{eq:Hhf}
H_\mathrm{hf} \simeq h_z S_z,~~ h_z = \sum_{k} A_k I_k^z.
\end{equation}
We set $\hbar = 1$, $\mathbf{h} = \sum_{k} A_k \mathbf{I}_k$ is the Overhauser operator, $\mathbf{S}=\boldsymbol{\sigma}/2$ is a pseudospin-$1/2$ operator in the two-dimensional ($J^z=\pm 3/2$) heavy-hole subspace, and $\mathbf{I}_k$ is the nuclear spin at site $k$. $A_k$ is the hyperfine coupling constant for the $k^{th}$ nuclear site and is given by $A_k = A^{i_k} v_0 |\psi(\mathbf{r}_k)|^2$, with $A^{i_k}$ the hyperfine constant for isotope $i_k$ at site $k$, $v_0$ the volume occupied by a single nuclear spin, and $\psi(\mathbf{r}_k)$ the heavy-hole envelope wavefunction~\cite{fischer2008spin,coish2009nuclear}. 
\begin{figure}[h]
\begin{centering}
\includegraphics[width = 0.5\textwidth]{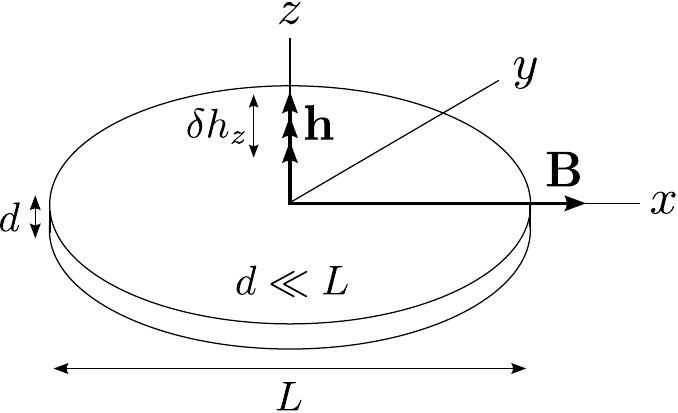}
\caption{Quantum-dot geometry, with Overhauser field $\mathbf{h}$ and magnetic field ${\bf B}=B\hat{x}$. For unstrained and flat ($d \ll L$) quantum dots, the in-plane hole gyromagnetic ratio is small, $g_\perp\simeq 0$~\cite{Korn2010}, and the Overhauser field lies along the [001] growth direction $\hat{z}$, $h_{x,y}\simeq 0$~\cite{fischer2008spin,Fischer2010,Maier2012}.}
\label{fig:geometry}
\end{centering}
\end{figure}

We consider an in-plane applied magnetic field $\mathbf{B} = B\hat{x}$ (see Fig. \ref{fig:geometry}), which results in the total Hamiltonian
\begin{equation}
\label{eq:fullH}
H = H_\mathrm{Z}+ H_\mathrm{hf},\quad H_\mathrm{Z}=-\gamma_\mathrm{H} B S_x - \displaystyle\sum\limits_{k} \gamma_k B I_k^x, 
\end{equation}
where $H_\mathrm{Z}$ gives the hole- and nuclear-Zeeman interactions, and $\gamma_k$ is the gyromagnetic ratio of the isotope at site $k$ having total spin $I_k$. For an InGaAs quantum dot, we take into account two isotopes of Ga (${}^{69}$Ga and ${}^{71}$Ga) with significantly different $\gamma$, and assume the same $\gamma$ for the two In isotopes (${}^{113}$In and ${}^{115}$In). The hole gyromagnetic ratio is $\gamma_\mathrm{H} = g_\bot \mu_\mathrm{B}$, with $g_\bot$ the in-plane $g$-factor for a dot with growth axis along [001] and $\mu_\mathrm{B}$ the Bohr magneton.

\begin{figure}
\begin{centering}
\includegraphics[width = 0.9\textwidth]{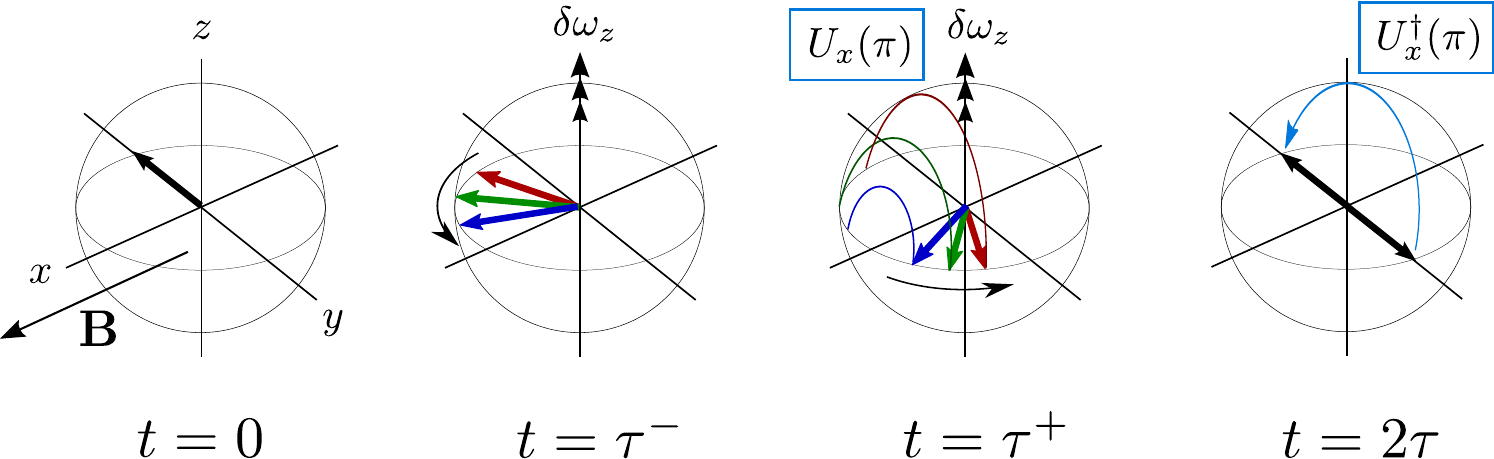}
\caption{Echo sequence with $\pi$-rotations about $\hat{x}$, which reverse dephasing from static fluctuations in an effective field $\delta\omega_z$ along $\hat{z}$ due to the hyperfine interaction. The magnetic field is $\mathbf{B} = B\hat{x}$ and in this figure we assume $g_\perp\simeq 0$ for simplicity such that there is no precession of the hole spin about $\hat{x}$.}
\label{fig:echo}
\end{centering}
\end{figure}

As the hole spin evolves under the action of the Hamiltonian in Eq.~(\ref{eq:fullH}), fluctuations in the Overhauser field $h_z$ induce fast spin dephasing.  This dephasing can be reversed with a spin echo if the fluctuations are effectively static on the timescale of spin decoherence. We consider an echo scheme consisting of $\pi$-pulses about $\hat{x}$, which would reverse the effects of static fluctuations in effective fields along $\hat{y}$ or $\hat{z}$, due to, e.g., the hyperfine interaction in Eq.~(\ref{eq:Hhf}) (see, e.g., Fig.~\ref{fig:echo}).

To calculate the effects of spin echo on hole-spin decay, we move into the interaction picture with respect to $H_\mathrm{Z}$ and define the interaction picture Hamiltonian $\tilde{H}(t)$ as
\begin{align}\label{eq:H_interaction}
\tilde{H}(t) &= e^{iH_\mathrm{Z} t}H_\mathrm{hf}e^{-iH_\mathrm{Z} t} = \tilde{h}_z(t)\tilde{S}_z(t),\\[6pt]
\label{eq:hz-interaction}
\text{with} ~~ \tilde{h}_z(t) &= \sum_k A_k[I_k^z \cos{(\gamma_{i_k} B t)} - I_k^y\sin{(\gamma_{i_k} B t)}] \\ 
\label{eq:Sz-interaction}
\text{and} ~~ \tilde{S}_z(t) &= [S_z\cos (\gamma_\mathrm{H} B t) - S_y\sin (\gamma_\mathrm{H} B t)]. 
\end{align}

Accounting for a $\pi$-pulse at time $\tau$ about an axis $\hat{\alpha}$ ($\alpha=x,y,z$), followed by a second $\pi$-pulse to return the spin to its original orientation, as illustrated in Fig.~\ref{fig:echo} for $\alpha=x$, the echo Hamiltonian is given by
\begin{align}
\tilde{H}_\mathrm{e}(t) &= \begin{cases}
\tilde{H}(t) &~~~~ 0\leq t < \tau,\\
\sigma_\alpha \tilde{H}(t) \sigma_\alpha &~~~~ \tau\leq t \leq 2\tau.
\label{eq:HIecho}
\end{cases}
\end{align}
From Eqs.~(\ref{eq:H_interaction}) - (\ref{eq:HIecho}), we see that if $\alpha = x$, then $\tilde{H}_\mathrm{e}(t) = -\tilde{H}(t)$ from $t=\tau$ to $t=2\tau$. Provided $\tilde{H}(t)$ is approximately static on this timescale, the $\pi$-pulse will reverse the evolution undergone by the system up until $t=\tau$ and refocus the hole spin signal at $t=2\tau$. Due to the anisotropy of the heavy-hole hyperfine interaction, the decay process is highly dependent on the chosen geometry; e.g. if $\alpha = y$ or $\alpha = z$, the echo does not reverse the sign of $\tilde{H}(t)$ from $t=\tau$ to $t=2\tau$. For this reason, we will focus on the case with the magnetic field along $\hat{x}$ and $\pi$-pulses performed about $\hat{x}$.

\subsection{Hole spin decoherence}

The hole-spin coherence dynamics can be written as
\begin{equation}
\big\langle S_{\alpha}(2\tau)\big\rangle = \big\langle \tilde{U}^{\dagger}(2\tau) \tilde{S}_{\alpha}(2\tau) \tilde{U}(2\tau) \big\rangle,
\end{equation}
where $\tilde{\mathcal{O}}$ denotes the operator $\mathcal{O}$ in the interaction picture. $\tilde{U}(2\tau)$, the time-evolution operator for $\pi$-pulses applied at times $t = \tau,2\tau$, is given by
\begin{equation}\label{eq:U}
\tilde{U}(2\tau) = \mathcal{T} e^{-i\int_0^{2\tau} dt \tilde{H}_\mathrm{e}(t)},
\end{equation}
with $\mathcal{T}$ the usual time-ordering operator. The Magnus expansion, an average-Hamiltonian theory typically applied to periodic and rapidly oscillating systems~\cite{Maricq1982}, allows us to rewrite the time-evolution operator as a series expansion,
\begin{equation}
\tilde{U}(2\tau) = e^{-iH_\mathrm{M}(2\tau)} = e^{-i\sum_{i=0}^\infty H^{(i)}(2\tau)},
\end{equation}
where all orders can be found by expanding Eq.~(\ref{eq:U}) in a Dyson series and applying the Baker-Campbell-Hausdorff formula. The first few terms are
\begin{align}
H^{(0)}(t) &= \int_0^t H(t_1) dt_1 \label{eq:H0-definition},\\[4pt]
H^{(1)}(t) &= -\frac{i}{2}\int_0^t \int_0^{t_2} \left[H(t_2),H(t_1)\right] dt_1dt_2 \label{eq:H1-definition}, \\[4pt]
H^{(2)}(t) &= -\frac{1}{6}\int_0^t \int_0^{t_3} \int_0^{t_2} ( [H(t_3),[H(t_2),H(t_1)]] + [H(t_1),[H(t_2),H(t_3)]] ) dt_1dt_2dt_3.
\end{align}
Each higher-order term in the Magnus expansion contains one additional integral over time, such that higher-order oscillating terms are suppressed by a factor of order $\| \tilde{H} \|/\omega$, with $\omega$ the typical oscillation frequency. In addition to bounded oscillatory terms, higher orders in the Magnus expansion typically also contain terms that grow in time.  To guarantee that these growing terms do not cause a significant modification of the dynamics evaluated here, we can only apply the Magnus expansion up to a timescale where these contributions reach $\sim 1$. The precise conditions for validity of the Magnus expansion in this case are given in Ref.~\cite{Wang2012} (see also the discussion in Ref.~ \cite{beaudoin2013enhanced}, which shows another application of the Magnus expansion technique to electron-spin dephasing).

We can now express the time-evolved spin components $S_\alpha (2\tau)$, where $\alpha = x,y,z$, as
\begin{equation}\label{eq:Liouvillian}
\big\langle S_{\alpha}(2\tau)\big\rangle
= 
\big\langle \left( e^{iL_\mathrm{M}(2\tau)} \tilde{S}_{\alpha}(2\tau)  \right) \big\rangle
\end{equation}
by defining the Liouvillian $L_\mathrm{M}(t)$~\cite{Wang2012} as
\begin{equation}
L_\mathrm{M}(2\tau)\mathcal{O} = \big[H_\mathrm{M}(2\tau), \mathcal{O}\big].
\end{equation} 
Here, the expectation value of a given operator $\mathcal{O}$ is defined as $\langle\mathcal{O}\rangle ={\rm Tr}\{\mathcal{O} \rho\}$. We make the assumption that the initial state $\rho = \rho_S \otimes \rho_I$ is a tensor product of the hole-spin ($\rho_S$) and nuclear-spin ($\rho_I$) density matrices. We further assume that the nuclear spins are in an infinite-temperature thermal state.  In this limit, for $N\gg 1$ uncorrelated nuclear spins, the central-limit theorem allows for the Gaussian approximation:
\begin{equation}\label{eq:momexp}
\big\langle e^{iL_\mathrm{M}(2\tau)} \tilde{S}_{\alpha} \big\rangle \simeq \bigg\langle \text{exp}\Big\{-\frac{1}{2}\big\langle L_\mathrm{M}^2(2\tau)\big\rangle_I \Big\} \tilde{S}_{\alpha}\bigg\rangle_S, 
\end{equation}
where, for any operator $\mathcal{O}_S$ acting in the hole-spin space,
\begin{equation}
\langle L_\mathrm{M}^2(t)\rangle_I\mathcal{O}_S = {\rm Tr}_I\{(L_\mathrm{M}^2(t)\mathcal{O}_S) \rho_I \}.
\end{equation}
In a sufficiently large magnetic field $B$, rapid oscillations in $\tilde{H}(t)$ will allow us to truncate the Magnus expansion at leading order, giving
\begin{equation}
L_\mathrm{M}(2\tau)\mathcal{O}_S \simeq L^{(0)}(2\tau)\mathcal{O}_S = [H^{(0)}(2\tau),\mathcal{O}_S]. 
\end{equation}

\subsection{Vanishing $g_\bot$ limit}

\begin{figure}
\begin{centering}
\includegraphics[width = 0.8\textwidth]{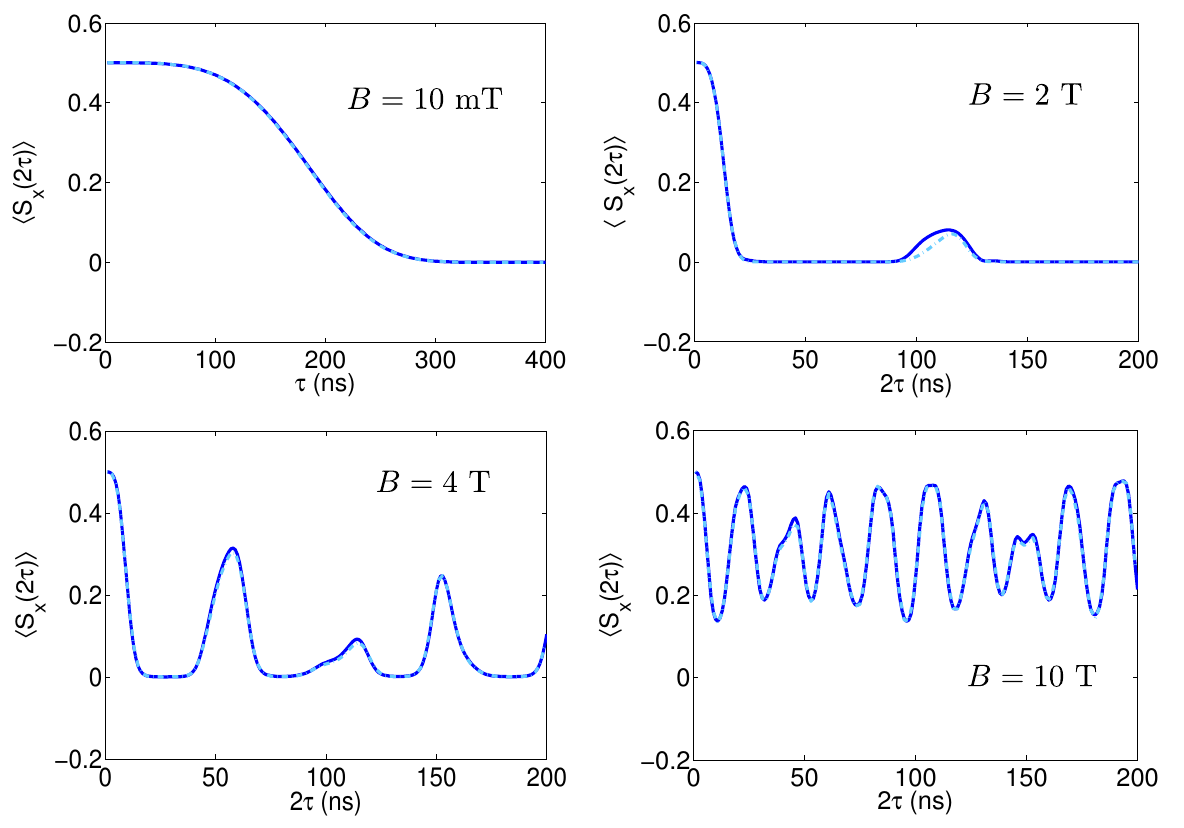}
\caption{Exact (solid lines) and approximate [dotted lines, from Eq.~(\ref{eq:magnus})] solutions for various magnetic field values $B$. Parameters used for these plots were: $\langle S_x(0) \rangle = 1/2, N = 10^4, \gamma_\mathrm{H} = g_\bot \mu_\mathrm{B} = 0$, and values for $\gamma_i$ can be found in Table I of Ref.~\cite{coish2009nuclear}.  We have further assumed isotopic abundances and nuclear spin appropriate for a $\mathrm{In}_x\mathrm{Ga}_{1-x}\mathrm{As}$ quantum dot with indium doping $x=0.5$, as given in Ref.~\cite{Wang2012}. }
\label{fig:Sx-exact}
\end{centering}
\end{figure}

In the limit $\gamma_\mathrm{H} = g_\bot\mu_\mathrm{B} = 0$, an exact analytical solution can be found for $\left<S_\alpha(t)\right>$  \cite{Wang2012,beaudoin2013enhanced}.  This exact solution can be used to verify the validity of approximations applied here and will provide a good description of the spin-echo dynamics whenever $\gamma_\mathrm{H} < \gamma_i$.  In GaAs, this condition corresponds to $g_\perp\lesssim 10^{-3}$ ($g_\perp<5\times 10^{-3}$ has been reported in 2D wells~\cite{Korn2010}). The exact evolution of $\langle S_x(2\tau) \rangle$ is shown in Fig.~\ref{fig:Sx-exact} for a range of $B$, revealing the transition between two distinct regimes: A low-$B$ regime resulting in complete decay, and a high-$B$ motional-averaging regime in which decay is bounded. In the low-$B$ regime (top two panels), the decay time decreases with increasing magnetic field, whereas for high $B$, the hole-spin coherence improves as we find periodic complete refocusing of the spin signal (bottom right panel). 

To gain physical intuition for this surprising behavior, we calculate the approximate dynamics by setting $\gamma_\mathrm{H}=0$ in Eq.~(\ref{eq:momexp}) and keep only the leading term in the Magnus expansion, giving~\cite{Wang2012}
\begin{equation}\label{eq:magnus}
\frac{\langle S_x(2\tau) \rangle} {\langle S_x(0) \rangle}\simeq
\exp{\left[ -\sum_i \frac{4\nu_i A^2 I_i(I_i+1)}{3 N (\gamma_i B)^2} \sin^4{\left(\frac{\gamma_i B \tau}{2}\right)}\right]},
\end{equation}
where the index $i$ sums over the different nuclear species, $\nu_i$ is the abundance of the $i^\text{th}$ species, and $A = \sum_k A_k$ is the total hyperfine coupling constant averaged over the nuclear species. Equation~(\ref{eq:magnus}) is plotted in Fig.~\ref{fig:Sx-exact} along with the exact solution. From this figure, we see that the approximate solution agrees almost perfectly with the exact solution in the chosen regime. 

In the low-$B$ regime, where $ B \ll A/(\gamma_i\sqrt{N})$ (top-left panel in Fig.~\ref{fig:Sx-exact}), a short-time expansion of Eq.~(\ref{eq:magnus}) gives 
\begin{align}
\langle S_x(2\tau) \rangle \simeq \langle S_x(0) \rangle \left(1-(\tau /\tau_0)^4\right) \simeq \langle S_x(0) \rangle e^{-(\tau /\tau_0)^4}, \\
\label{eq:tau_0}
\text{with}~~\tau_0 \simeq \frac{1}{\sqrt{B}}\left[\sum_i \frac{\nu_i ( \gamma_i A)^2}{4N} \frac{I_i(I_i+1)}{3}\right]^{-1/4}.
\end{align}
When $B$ increases, Eq.~(\ref{eq:tau_0}) shows that the initial decay time $\tau_0$ decreases, in agreement with the top two panels of Fig.~\ref{fig:Sx-exact}. This behavior is contrary to the usual expectation for electron spins (for electrons, the coherence time normally increases with increasing $B$ \cite{Cywinski2009}).  Here, a decrease in the decay time with increasing $B$ is due to dynamical fluctuations in $h_z$ induced by rapidly precessing nuclear spins.  Because the nuclear-spin system fluctuates in time, decay due to this dynamical environment can no longer be completely reversed by a simple Hahn echo scheme. 

At large magnetic fields, when $B \gtrsim A/(\gamma_i \sqrt{N})$, we see from Eq.~(\ref{eq:magnus}) and Fig.~\ref{fig:Sx-exact} (bottom right panel) that the hole-spin signal no longer decays to zero. In fact, according to Eq.~(\ref{eq:magnus}), the oscillation amplitude in the exponent decreases as $1/B^2$. In this regime, the nuclear spins precess rapidly enough to average out the fluctuations in $h_z$; this motional averaging accounts for the bounded decay as well as the suppression in oscillation amplitude of the echo signal. Between the two regimes, we find partial periodic recovery of the hole-spin signal.

While the discussion above addresses the $g_\bot = 0$ case, we predict the same motional averaging effect to occur when $\gamma_\mathrm{H} \gg \gamma_i$, at even lower applied field values~\cite{Wang2012}. The main qualitative distinction from the $g_\bot = 0$ case is an additional beating in the echo envelope function introduced by the hole Zeeman frequency. These results demonstrate that hole-spin dephasing due to hyperfine couplings can be completely suppressed at moderate magnetic fields, in contrast with the case of electron spins.

To conclude this section, we mention a number of other interesting topics related to hole-spin coherence and manipulation. First, the spin-echo decay observed in Ref.~\cite{DeGreve2011} is believed to be primarily caused by electric-field-induced noise, instead of hyperfine coupling to nuclear spins. In light of this result, in Ref.~\cite{Wang2012} we have developed a simple phenomenological theory of spin decay brought on by fluctuations in the hole-spin precession frequency. Furthermore, new schemes for potentially improved single-hole spin manipulation and control have been proposed very recently as well. One such method suggests coherently manipulating spin states through the Dresselhaus spin-orbit interaction, where the heavy-hole spin would be transported around closed loops using static applied voltages~\cite{Szumniak2012}. Another proposal points to the fact that a hole spin's precession can be altered through electric-field-induced g-tensor modulations, resulting in faster predicted spin manipulation times for holes than for electrons~\cite{Pingenot2010}. A quantum-dot molecule has also been proposed as a potentially scalable qubit architecture which could allow for enhanced wavelength tunability and qubit rotation fidelity~\cite{Economou2012,Muller2012,Gawarecki2012,Greilich2012arxiv}. In addition to the latest advances in maintaining and controlling the hole spin, current and future prospects are paving the way for establishing hole spins in quantum dots as a robust qubit implementation. 

\section{Conclusions and open questions}\label{Conclusions}

In this Chapter we have reviewed recent work on heavy holes in III-V semiconductors, by focusing on many-body phenomena due to electron-interactions (for two-dimensional systems) and the hyperfine interactions with the nuclear bath (for single-hole quantum dots). In both cases, we have tried to highlight the unique behavior of heavy holes, which originates from the total angular momentum $J_z=\pm 3/2$ of the effective spin.  For two-dimensional liquids we have discussed how typical non-linear-in-momentum spin-orbit couplings for holes can have a stronger interplay with the Coulomb repulsion. In particular, the significant difference in group-velocities for holes with strong spin-orbit coupling implies that the spatial separation of spin components is much easier to achieve than for electron carriers. This velocity difference could be observed with optical techniques including Raman scattering and direct time-resolved Faraday-rotation detection of hole-wavepacket motion, in addition to transport measurements of the effective mass. 

Other properties of the hole liquid will likely show similarly interesting behavior. Future studies might include a complete characterization of the quasiparticle parameters (lifetime, spectral weight) and compressibility.  The challenging regime of dilute hole liquids with large interaction parameters $r_s \simeq 6 - 12$ calls for further progress in non-perturbative approaches, e.g., a suitable extensions of the Jastrow-factor variational wavefunctions and the Quantum Monte Carlo method. 

On the other hand, single holes in quantum dots represent an emerging platform for the implementation of spin qubits and are attracting a growing interest, as is clear from the flourishing experimental progress of the field. Following the first realization of the Hahn spin echo in this system \cite{DeGreve2011}, we have described a theoretical analysis of spin-echo decay based on exact results and a systematic method of solution in terms of a Magnus expansion \cite{Wang2012}. We expect this approach will be a useful tool for further studies of spin decoherence, in this or other related systems. For example, spin-dephasing due to the hyperfine interaction in the presence of a magnetic field gradient was analyzed in Ref.~\cite{beaudoin2013enhanced} through an exact solution and an approximate Magnus expansion analogous to the ones discussed here. While electric noise seems to limit the spin-echo decay in current experiments on holes \cite{DeGreve2011}, the prediction of a motional-averaging regime at high magnetic field will be especially interesting for a future generation of devices, dominated by hyperfine decoherence.  The presence of such a regime strengthens the case that holes will be able to realize long-lived spin qubits, amenable to fast optical manipulation. 

\acknowledgments
We acknowledge financial support from NSERC, CIFAR, FRQNT, and INTRIQ.

\bibliographystyle{varenna}
\bibliography{VarennaBibliography}

\end{document}